\documentclass[twocolumn]{aastex631}

\usepackage{microtype}
\usepackage{multirow}
\usepackage{amsmath}

\begin{document}

\title{PSR J1922+37: a 1.9-second pulsar discovered in the direction of the old open cluster NGC 6791}

\author[0000-0002-2187-4087]{Xiao-Jin Liu}
\affiliation{Department of Physics, Faculty of Arts and Sciences, Beijing Normal University, Zhuhai 519087, China}
\affiliation{Advanced Institute of Natural Sciences, Beijing Normal University, Zhuhai 519087, China}

\author[0000-0002-9409-3214]{Rahul Sengar}
\affiliation{Max Planck Institute for Gravitational Physics (Albert Einstein Institute), D-30167 Hannover, Germany}
\affiliation{Leibniz Universität Hannover, D-30167 Hannover, Germany}

\author[0000-0003-3294-3081]{Matthew Bailes}
\affiliation{Centre for Astrophysics and Supercomputing, Swinburne University of Technology, P.O. Box 218, Hawthorn, VIC 3122, Australia}
\affiliation{OzGrav: The ARC Centre of Excellence for Gravitational Wave Discovery, Hawthorn, VIC 3122, Australia}

\author[0000-0001-6196-4135]{Ralph P. Eatough}
\affiliation{CAS Key Laboratory of FAST, National Astronomical Observatories, Chinese Academy of Sciences, Beijing, 100101, China}
\affiliation{Max-Planck-Institut f\"ur Radioastronomie, Auf dem H\"ugel 69, D-53121 Bonn, Germany}

\author[0000-0002-5381-6498]{Jianping Yuan}
\affiliation{Xinjiang Astronomical Observatory, Chinese Academy of Sciences, 150 Science 1-Street, Urumqi, Xinjiang 830011, China}

\author[0000-0002-9786-8548]{Na Wang}
\affiliation{Xinjiang Astronomical Observatory, Chinese Academy of Sciences, 150 Science 1-Street, Urumqi, Xinjiang 830011, China}

\author[0000-0001-5105-4058]{Weiwei Zhu}
\affiliation{CAS Key Laboratory of FAST, National Astronomical Observatories, Chinese Academy of Sciences, Beijing, 100101, China}
\affiliation{Institute for Frontier in Astronomy and Astrophysics, Beijing Normal University, Beijing 102206, China}

\author[0009-0000-1929-7121]{Lu Zhou}
\affiliation{School of Physics and Technology, Wuhan University, Wuhan, Hubei 430072, China}
\affiliation{Advanced Institute of Natural Sciences, Beijing Normal University, Zhuhai 519087, China}

\author[0000-0003-2516-6288]{He Gao}
\affiliation{School of Physics and Astronomy, Beijing Normal University, Beijing 100875, China}
\affiliation{Institute for Frontier in Astronomy and Astrophysics, Beijing Normal University, Beijing 102206, China}

\author[0000-0002-3567-6743]{Zong-Hong Zhu}
\affiliation{School of Physics and Astronomy, Beijing Normal University, Beijing 100875, China}
\affiliation{Institute for Frontier in Astronomy and Astrophysics, Beijing Normal University, Beijing 102206, China}
\affiliation{School of Physics and Technology, Wuhan University, Wuhan, Hubei 430072, China}

\author[0000-0001-7049-6468]{Xing-Jiang Zhu}
\correspondingauthor{Xing-Jiang Zhu}
\email{zhuxj@bnu.edu.cn}
\affiliation{Department of Physics, Faculty of Arts and Sciences, Beijing Normal University, Zhuhai 519087, China}
\affiliation{Institute for Frontier in Astronomy and Astrophysics, Beijing Normal University, Beijing 102206, China}
\affiliation{Advanced Institute of Natural Sciences, Beijing Normal University, Zhuhai 519087, China}

\begin{abstract}

More than 300 pulsars have been discovered in Galactic globular clusters; however, none have been found in open clusters. Here we present results from a 20-hour survey of seven open clusters with the Five-hundred-meter Aperture Spherical radio Telescope (FAST).
Our first discovery is a 1.9-second pulsar (J1922+37) found in the direction of the old open cluster NGC 6791.
The measured dispersion measure (DM) implies a distance of 4.79 kpc and 8.92 kpc based on the NE2001 and YMW16 electron density models, respectively. Given the large uncertainty of DM distance estimates, it is plausible that PSR J1922+37 is indeed a member of NGC 6791, for which the distance is $4.19\pm0.02$ kpc based on Gaia Data Release 3.
If confirmed, PSR J1922+37 will be the first pulsar found in a Galactic open cluster.
We outline future observations that can confirm this pulsar-open cluster association and discuss the importance of this pulsar
for calibrating the characteristic ages
of the pulsar population.

\end{abstract}

\keywords{Radio pulsars (1353) --- Open star clusters (1160) --- Globular star clusters (656)}
 
\section{Introduction} \label{sec:intro}

Pulsars are well known for their fast spin, high density, and strong gravitational and magnetic field, rendering them astrophysical laboratories unachievable on Earth \citep[see e.g.][]{LK12}. 
Globular clusters have the highest spatial concentrations of pulsars, with more than 300 pulsars discovered in 45 globular clusters\footnote{\url{https://www3.mpifr-bonn.mpg.de/staff/pfreire/GCpsr.html}}.
In particular, 47 Tucanae alone has more than 40 pulsars, while Terzan 5 has nearly 50 pulsars. 
The large number of pulsars has provided a unique tool to probe the density profile and the dynamics of their host clusters, and to constrain the mass of potential intermediate massive black holes inside \citep{FRK+17, PRF+17, PSL+17}.
The high yield of pulsars in globular clusters is believed to be caused by their high stellar encounter rate, which is directly related to their crowded stellar environments \citep{VER02, HCT10, BHS+13, VF14}.

Open clusters also have a relatively high stellar density, and more than 2900 such clusters have been identified, about 10 times the number of globular clusters \citep[e.g.][and their Table 1]{BPB+19}. 
However, no radio pulsars have been discovered in any open clusters yet\footnote{An X-ray pulsar and magnetar, CXO~J164710.2$-$455216, was discovered in the Galactic young massive cluster Westerlund~1 \citep{MCC+06}.  Based on epicyclic approximation and simulations, PSR~1932+1059 was proposed to be a runaway system from the open cluster IC~4665 \citep{BOB08}.}. 
The difference is probably due to their smaller cluster mass, which is usually several orders of magnitude smaller than that of globular clusters.
Considering the total number of open clusters and globular clusters and
assuming a similar formation rate of neutron star per unit mass, this means fewer neutron stars could be present in open clusters.
In addition, the lower escape velocity of open clusters also makes it more difficult to retain pulsars that usually suffer from natal kicks \citep[e.g.][]{PRP02, IHR+08, BLT+11}.
The short lifetime ($\sim 10^8$ yr) of open clusters also means that they are likely to have disintegrated before any pulsars were discovered inside. 
As a result, open clusters were unfavorable places for pulsar searching and were usually overlooked in past search efforts.
However, in the search for unusual types of pulsar, open clusters deserve more attention.
Any confirmed pulsar discovery in these clusters will greatly advance our understanding of stellar evolution and pulsar formation. It would tell us that some pulsars can be born with very small kicks, and also permit a comparison of the pulsar's characteristic age with that of the open cluster.

As the largest single-dish telescope, the Five-hundred-meter Aperture Spherical radio Telescope (FAST) \citep{NLJ+11} provides an unprecedented sensitivity for pulsar searches. 
In addition, the adoption of a 19-beam receiver also provides a decent survey speed to observe a large area \citep{JTH+20}. 
Thus, FAST has been proven to be a productive telescope for pulsar discovery. 
To date, FAST has discovered more than 1000 new pulsars in the past few years, mainly through the Galactic Plane Pulsar Survey (GPPS, \citealt{HWW+21}), the Commensal Radio Astronomy FAST Survey (CRAFTS, \citealt{LWQ+18}) and dedicated globular cluster searches \citep[e.g.][]{PQM+21}. 

In this paper, we present results from a 20-h FAST observing project (ID: PT2021\_0102), with the goal of finding exotic pulsar systems in seven open clusters. 
After initial analysis of these observations, we found a 1.9-second pulsar, J1922+37, which may be associated with the open cluster NGC~6791.
If confirmed, J1922+37 will be the first pulsar discovered in an open cluster and will provide new insights into stellar evolution and neutron star production in open clusters.

The remainder of the paper is organized as follows. 
We first describe the observations in Section~\ref{sec:obs}, then introduce the search method in Section~\ref{sec:method} and present our results in Section~\ref{sec:result}, before discussing the results in Section~\ref{sec:discussion}. 
The paper is concluded in Section~\ref{sec:summary}.

\section{Observations} \label{sec:obs}

Using the Milky Way star cluster catalog\footnote{\url{https://heasarc.gsfc.nasa.gov/W3Browse/star-catalog/mwsc.html}} \citep{KPS+13}, we chose seven open clusters that are old and massive as targets. 
The basic information of the clusters is shown in Table~\ref{tab:oc_obs}.
We see that all clusters are within $\sim10^\circ$ of the Galactic plane. 
If no pulsars are found in the open clusters, we may still be able to detect pulsars in the foreground or background sky; note that the same sky locations (of these open clusters) have been or will be covered by the GPPS survey \citep{HWW+21}, which motivated us to adopt a longer integration time for our observations (see Tables \ref{tab:obs_mode} and \ref{tab:obs_mode_snapshot} for details).

Observations were carried out using all beams of the L-band 19-beam receiver \citep{JTH+20} with a time resolution of 49.152~$\mu$s.
Two polarizations and 4096 frequency channels were used throughout the observations.
The receiver provides a bandwidth of 500~MHz, but only 400~MHz are used in pulsar searches. 
Data were recorded with the pulsar backend in search mode and saved in psrfits format \citep{HVM04} with a default bit depth of 8.
For calibration purposes, each observation began with a low-intensity noise injection, which has a modulating period of 0.2~second and lasted for one minute.

Taking into account the gaps between individual beams, four pointings are necessary to cover a given region \citep[e.g.][]{JTH+20}. 
For efficient coverage, all clusters were observed using two groups of on-off mode observations\footnote{\url{https://fast.bao.ac.cn/cms/article/24/}}, except for IC~4756, which was observed with the more efficient snapshot mode, the same observation mode used for the GPPS survey \citep{HWW+21}.
Due to the large angular size of IC~4756, two groups of snapshot-mode observations were applied to cover the central region of the cluster. 
The observation position and integration length of the clusters are given in Tables~\ref{tab:obs_mode} and \ref{tab:obs_mode_snapshot} for the on-off mode and the snapshot mode, respectively.
For the on-off mode, we provide the position of the central beam (M01) of each pointing, while for the snapshot mode, only the central beam of the first pointing is provided. 
In both modes, the position of the remaining beams can be deduced by accounting for the geometry of the 19-beam receiver and considering the slewing track of the receiver in each mode \citep{JTH+20, HWW+21}.

\begin{table*}[]
    \centering
    \caption{Summary of the seven open clusters observed in this project. Columns are the cluster name, equatorial and Galactic coordinates, tidal radius, age, distance, and the estimated dispersion measure (DM) based on the distance and the NE2001 and YMW16 electron density model. The cluster parameters have been updated using \cite{HR24}.}
    \begin{tabular}{cccccccccc}
        \hline\hline
         Open cluster & RA  & DEC & Gl & Gb & $R$ & Age & $D$ & DM$_{\rm NE2001}$ & DM$_{\rm YMW16}$ \\
         & (hhmmss) & (ddmmss) & (deg) & (deg) & (arcmin) & ($10^8$~yr) & (kpc) & (cm$^{-3}$~pc) & (cm$^{-3}$~pc) \\
         \hline
    NGC 1912/M38 & 05:28:40  &  +35:49:15  &  172.27   &  0.68  &  40.4   &  2.7   &   1.09  &   34.8   &   38.5      \\
    NGC 2099/M37 & 05:52:19  &  +32:32:27  &  177.65   &  3.09  &  30.4   &  4.9   &   1.40  &   46.5   &   60.3      \\
    NGC 2158 & 06:07:26  &  +24:05:52  &  186.64   &  1.78  &  12.6   &  18.6  &   3.71  &   128.2  &   172.8     \\
    NGC 2168/M35 & 06:08:48  &  +24:17:11  &  186.62   &  2.15  &  14.3   &  6.3   &   2.40  &   91.2   &   149.5     \\
    IC 4756  & 18:38:31  &  +05:26:14  &  36.31    &  5.34  &  94.6   &  8.4   &   0.47  &   4.8    &   7.3       \\
    NGC 6791 & 19:20:53  &  +37:46:27  &  69.96    &  10.91 &  11.6   &  38.0  &   4.19  &   74.1   &   49.6      \\
    NGC 7789 & 23:57:19  &  +56:43:26  &  115.52   &  $-5.37$ &  32.2   &  15.7  &   1.97  &   44.2   &   47.9      \\
         \hline
    \end{tabular}
    \label{tab:oc_obs}
\end{table*}

\begin{table*}
    \centering
    \caption{The observation position, integration length and observation date of open clusters observed with the on-off mode. Each cluster was observed with two groups of on-off mode hence four pointings. For each pointing, the position of the central beam (M01) is listed. $t_{\rm int}$ is the approximate integration length of each pointing, while the observation date is based on UTC+8.}
    \begin{tabular}{ccccccc}
        % \\
        \hline\hline
        Open cluster & on RA & on DEC & off RA & off DEC & $t_{\rm int}$ & Obs. Date\\
        & (hhmmss) & (ddmmss) & (hhmmss) & (ddmmss) & (min) & (yy-mm-dd) \\
        \hline
        \multirow{2}{*}{NGC 1912/M38} & 05:28:52.00 & +35:48:00.0 & 05:29:06.80 & +35:48:00.0 & \multirow{2}{*}{27}  & \multirow{2}{*}{2021-09-27} \\
        & 05:28:44.60 & +35:50:36.0 & 05:28:59.40 & +35:50:36.0 \\
        \hline
        \multirow{2}{*}{NGC 2099/M37} & 05:52:21.00 & +32:34:12.0 & 05:52:35.24 & +32:34:12.0  & \multirow{2}{*}{27}  & \multirow{2}{*}{2021-09-28}\\ 
        & 05:52:13.88 & +32:36:48.0 & 05:52:28.12 & +32:36:48.0 \\
        \hline
        \multirow{2}{*}{NGC~2158}& 06:07:26.00 & +24:05:30.0 & 06:07:39.15 & +24:05:30.0 & \multirow{2}{*}{57}  & \multirow{2}{*}{2021-09-29}\\
        & 06:07:19.43 & +24:08:06.0 & 06:07:32.57 & +24:08:06.0 \\
        \hline
        \multirow{2}{*}{NGC 2168/M35} & 06:09:13.00 & +24:21:36.0 & 06:09:26.17 & +24:21:36.0 & \multirow{2}{*}{27} & \multirow{2}{*}{2021-09-30} \\
        & 06:09:06.41 & +24:24:12.0 & 06:09:19.59 & +24:24:12.0 \\
        \hline
        \multirow{2}{*}{NGC~6791} & 19:20:53.00 & +37:46:48.0 & 19:21:08.18 & +37:46:48.0 & \multirow{2}{*}{57} & \multirow{2}{*}{2022-01-22}\\
        & 19:20:45.41 & +37:49:24.0 & 19:21:00.59 & +37:49:24.0 \\
        \hline
        \multirow{2}{*}{NGC~7789} & 23:57:25.00 & +56:43:48.0 & 23:57:46.87 & +56:43:48.0 & \multirow{2}{*}{27} & \multirow{2}{*}{2021-09-29}\\
        & 23:57:14.06 & +56:46:24.0 & 23:57:35.94 & +56:46:24.0 \\
         \hline
    \end{tabular}
    \label{tab:obs_mode}
\end{table*}

\begin{table*}
    \centering
    \caption{The observation position, integration length and observation date of IC~4756. Since the snapshot mode was used for the cluster, only the position of the central beam (M01) in the first pointing is shown.}
    \begin{tabular}{ccccc}
        \hline \hline 
        Open cluster & RA & DEC  & $t_{\rm int}$ & Obs. Date\\
        & (hhmmss) & (ddmm) & (min)  & (yy-mm-dd) \\ 
        \hline
        \multirow{2}{*}{IC~4756 } & 18:39:38.19 & +05:29:36.0 & 18 & 2022-01-29\\
         & 18:38:01.75	& +05:29:36.0 & 28 & 2022-02-03\\
        \hline
    \end{tabular}
    \label{tab:obs_mode_snapshot}
\end{table*}

\section{Search method} 
\label{sec:method}

A bottleneck of processing pulsar search-mode data is the slow I/O speed due to the large data volumes. 
To reduce data size, we converted the data from 8 bits to 2 bits and halved the number of frequency channels using the \textit{digifil} routine in the \textsc{dspsr} software package \citep{vB11}. 
The conversion thus reduced the size of the data by a factor of 8, but only leads to a few percent decrease in signal-noise-ratio (S/N), as demonstrated using the known pulsar J1837+0528g\footnote{\url{http://zmtt.bao.ac.cn/GPPS/GPPSnewPSR.html}} in the line-of-sight of IC~4756 (see Section~\ref{sec:discovery}). 
After conversion, the data have 2048 frequency channels while the time resolution remains unchanged.

To minimize the impact of radio frequency interference (RFI), we found the bad frequency channels in each observation using the \textit{rfifind} routine in \textsc{presto} software package \citep{RAN11}. 
A length of one second was used to integrate for the \textit{rfifind} analyses. 
The bad frequency channels were masked out for the remaining processing.

We also removed the periodic interference using the \textsc{sigpyproc} software package\footnote{\url{https://sigpyproc3.readthedocs.io/en/latest/}}. 
For this purpose, we first removed the bad frequency channels from each observation. For each beam, a time series was then generated at zero DM and a Fast Fourier Transform (FFT) was performed to obtain the power spectrum using \textsc{sigpyproc}. Since the receiver's 19 beams observe the sky simultaneously, any periodic interference would appear across multiple beams. Therefore, to identify such periodic radio frequency interference or ``birdies", we flagged frequency bins if they exceeded a certain threshold ($\sigma>4$) in the power spectrum of four or more beams. These frequencies were used as a common birdie list for all 19-beams of a single pointing.

To de-disperse the data, we searched the range of DM in [2, 600]~cm$^{-3}$~pc, where the upper end is much larger than the largest DM value estimated using the cluster distance and the NE2001 \citep{CL02} or YMW16 \citep{YMW16} electron density model, see Table~\ref{tab:oc_obs}. 

To minimize the loss of sensitivity to pulsars in compact binaries \citep[e.g.][]{JK91, RCE+03}, we also searched in acceleration space. 
Considering the computational cost and the large number of candidates for a thorough search, we first conducted a fast search in a small acceleration range, $|a|\le1$~m$~$s$^{-2}$, while a deeper search considering a larger acceleration range is ongoing and the results will be presented elsewhere.

Both de-dispersion and acceleration searches were performed with the \textsc{peasoup}\footnote{\url{https://github.com/ewanbarr/peasoup}} software package, which can also take into account the bad frequency channels and birdies mentioned above.
More importantly, \textsc{peasoup} is based on the Graphic Processing Unit (GPU), which can significantly speed up the search process \citep[e.g.][]{MBC+19, SBB+23, SBB+24}.
For each search, in the spectral domain we set an S/N threshold of 6.5 and only kept the candidates above that threshold. As a result, about 15000 candidates were found for each pointing.

To reduce the number of candidates to a manageable scale, we adopted a parameter-based method, which uses a combination of the candidate period, S/N and the number of harmonics to pick out the most probable pulsar candidates as in \citet{SBB+23}.
We then folded the remaining candidates' search-mode data using the period, DM, and acceleration, with the \textit{dspsr} routine in the \textsc{psrchive}\footnote{\url{https://psrchive.sourceforge.net/}} software package \citep{HVM04}. 
The resulting archive files were first cleaned with the interference removal program \textsc{clfd}\footnote{\url{https://github.com/itachi-gf/clfd/tree/master}} \citep{MBC+23}, before being processed with the \textit{pdmp} routine in \textsc{psrchive}, which optimizes the period, DM, and S/N and provides diagnostic plots for human inspection.
Using the refined parameters, the sifting and folding process was repeated again, further reducing the number of candidates and creating diagnostic plots for these candidates, which were visually inspected to pick out any known pulsars and high significance candidates. 

\section{Results}
\label{sec:result}

\subsection{Discovery}
\label{sec:discovery}

Two pulsars were detected in the search, PSRs~J1837+05 and J1922+37.
PSR~J1837+05 is a known 6.26~ms pulsar and has a DM of $\sim120.8$~pc~cm$^{-3}$, consistent with PSR~J1837+0528g, which was discovered by GPPS.
J1837+05 was detected in three adjacent beams, which are also consistent with the discovery position of J1837+0528g, confirming the association with the known pulsar. 
J1837+05 is in the line-of-sight of open cluster IC~4756, but its DM is about 20 times larger than the DM estimate using the cluster distance and the electron density models (see Table~\ref{tab:oc_obs}).
J1837+05 is thus more likely in the background than in the open cluster.

\begin{figure*}
    \centering
    \includegraphics[width=1.6\columnwidth]{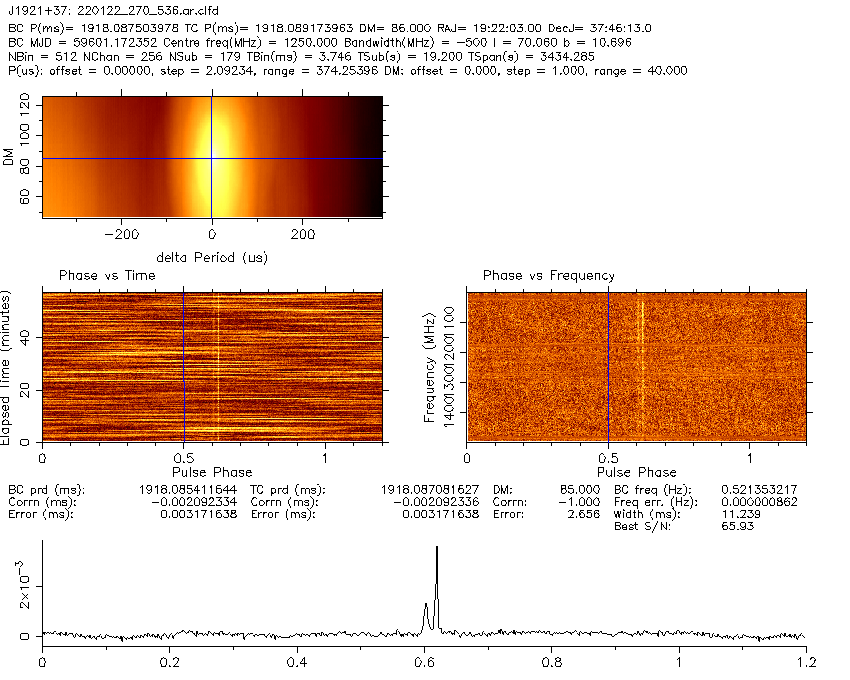}
    \caption{The discovery plot of PSR~J1922+37 from the \textit{pdmp} routine in \textsc{psrchive}. The top panel shows the optimization of period and DM, while the middle panels are phase-time and phase-frequency plots and the bottom panel shows the pulse profile.}
    \label{fig_J1922}
\end{figure*}

\begin{figure*}
    \centering
    \includegraphics[width=1.5\columnwidth]{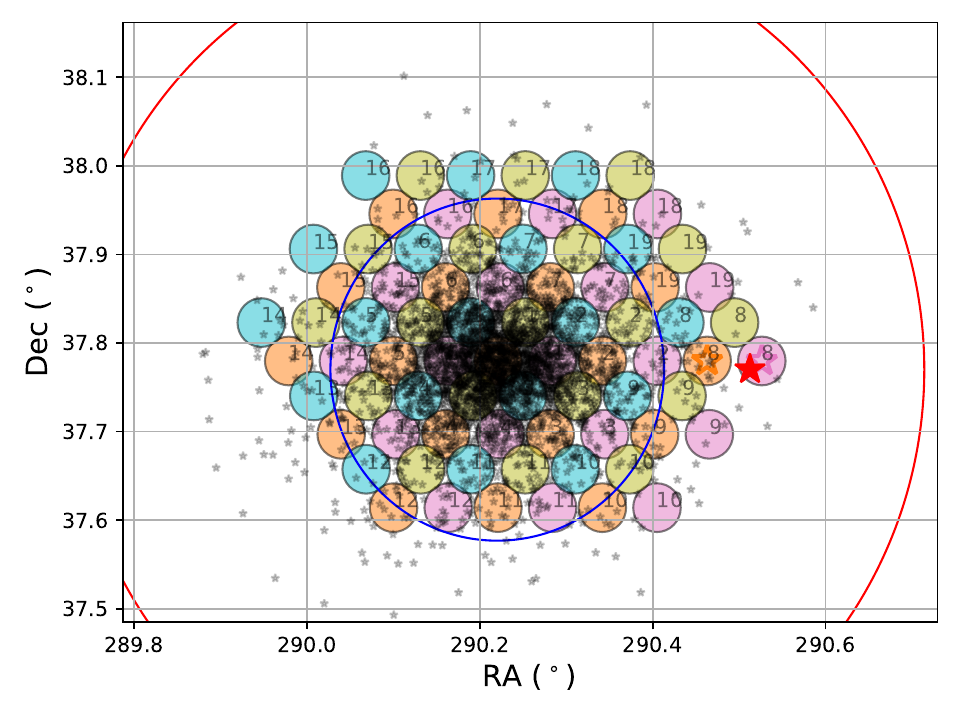}
    \caption{The position of J1922+37 and the member stars of NGC~6791. 
    The filled circles in orange, pink, cyan, and olive show the tiled beams of all four observation pointings. 
    The two discovery beams (M08) are indicated by the hollow stars, while the optimized position is labeled by the solid red star, and the cluster members are marked by the gray stars. The blue and red circles show the boundaries of the tidal and Jacobi radii respectively.}
    \label{fig_NGC6791}
\end{figure*}

PSR~J1922+37 has a period of 1.92~s and a DM of $\sim85$~pc~cm$^{-3}$ (Fig.~\ref{fig_J1922}), and is in the line-of-sight of open cluster NGC~6791 (Fig.~\ref{fig_NGC6791}).
The pulsar was discovered on 22 January 2022 in two adjacent beams (M08) with a comparable S/N $\sim60$.
In addition to the fundamental spin period, many harmonics were also detected.
No significant acceleration was detected.
Using FAST, the pulsar was later confirmed in an 18-min follow-up observation and in a 6-min gridded observation (see Section~\ref{sec:position_distance} for more details).

\subsection{Pulse width and flux density}

To estimate the flux density of PSR~J1922+37, we measured its pulse width. 
We fitted the pulse region with a two-component Gaussian model to account for both the main- and sub-pulse. 
As an approximation, we use the full width at half maximum (FWHM) of the main pulse as the pulse width, $0.012\pm0.002$~s, which gives a duty cycle of $0.63\%\pm0.08\%$. 

We estimate the flux density by the radiometer equation \citep{LK12}, 
\begin{equation}
    S = \frac{{\rm S/N}~T_{\rm sys}}{G\sqrt{n_{\rm p}t_{\rm int}\Delta f}}\sqrt{\frac{W}{P-W}},
\end{equation}
where we use $T_{\rm sys}=18$~K, $\Delta f=400$~MHz \citep{JYG+19}, and $G$ is the effective gain.
For the discovery observation, which has the highest S/N~=~66, $n_{\rm p}=2$ and $t_{\rm int}=3400$~s.
According to \cite{JTH+20}, the gain of beam M08 is about 0.92 times that of the central beam at 1250~MHz, which gives $G_0 = 14.8$~K~Jy$^{-1}$.
Note that the center of the discovery beams is about half a beam-width off the source, so the effective gain should be smaller.
If we assume a Gaussian beam pattern \citep[e.g.][]{CR16}, then $G = G_0/2$.
Using the parameters above, we obtain a flux density of just $\sim7~\mu$Jy.

The sparse number of observations mentioned above were spread over 3 years and are thus insufficient to produce a phase-coherent timing solution.
As a result, spin-down and characteristic age are unavailable at the moment.
No reliable polarization information was obtained, as all the observations either do not have sufficiently high S/N or full Stokes parameters, making the polarization measurement impossible. 

\subsection{Position and distance}
\label{sec:position_distance}

We refined the pulsar position using two groups of on-off-mode observations with FAST and the same receiver. 
The observations were centered on the midpoint of the two adjacent discovery beams and fully covered the nearby region in grids.
Each pointing was 6 minutes in duration.
The pulsar was detected in three adjacent beams, one beam reporting an S/N of $\sim17$ and two beams of $\sim7$. 
Using the S/N-weighted sum of the center of the beams, we updated the pulsar position to RA = 19:22:03 and Dec = +37:46:13 with a conservative uncertainty of the beam diameter ($\sim 3\arcmin$).

According to the position, PSR~J1922$+$37 is in the same direction as NGC~6791 with an offset of $\sim14\arcmin$ from the center of the cluster. Fig.~\ref{fig_NGC6791} shows the member stars of NGC~6791, which are taken from \cite{HR24} and have a membership probability greater than 30\%\footnote{\url{https://cdsarc.cds.unistra.fr/viz-bin/cat/J/A+A/686/A42\#/article}}.
NGC~6791 has a half-mass radius of $r_{\rm h}=4\farcm42$ \citep{PCK+11}.
PSR~J1922+37 is slightly outside the \cite{King62} tidal radius ($r_{\rm t}=11\farcm6$) of NGC~6791 but still within its Jacobi radius ($r_{\rm J}=29\farcm7$), an alternative to estimate the radius of gravitational influence for cluster retention \citep{BT08, HR24}.

Using the optimal DM and position, the NE2001 electron density model \citep{CL02} gives a distance estimate of $d_{\rm NE2001}=4.79\pm0.96$~kpc (assuming a nomial 20\% uncertainty, e.g. \citealt{SCL20}), which is close to the cluster distance, $d_{\rm NGC6791}$, reported by the Milky Way star cluster survey (4.93~kpc, \citealt{KPS+13}) and that inferred from Gaia Data Release 2 \citep[$4.45\pm0.10$~kpc,][]{DMM+21}, but slightly differs from that obtained using red clump giants in the cluster \citep[$4.02\pm0.15$~kpc][]{GL12} and that using Gaia Data Release 3 \citep[$4.19\pm0.02$~kpc,][]{HR24}.
However, the distance estimated by the YMW16 electron density model \citep{YMW16} is $d_{\rm YMW16}= 8.9\pm1.8$~kpc, about two times the distance to NGC~6791. Clearly we cannot definitively
argue for the association based solely on the DM. 
% Another possibility is to measure the distance through HI absorption line measurements 

\section{Discussion}
\label{sec:discussion}

The close vlaue between $d_{\rm NE2001}$ and $d_{\rm NGC6791}$ makes PSR~J1922+37 a plausible member at the edge of the NGC~6791. 
In fact, there are three Globular cluster pulsars that reside reasonably far from their corresponding cluster centers, including PSRs~B1718$-$19A \citep[NGC~6342,][]{LBH+93}, J1911$-$5958A \citep[NGC~6752,][]{DLM+01} and J1801$-$0857D \citep[NGC~6517,][]{LRF+11}.
A fourth pulsar, J1823$-$3022, is at $\sim3$ times the half-light radius of NGC~6624, although its association still awaits confirmation \citep{ARB+22}.

If our open cluster association is confirmed, PSR~J1922+37 will be the first radio pulsar found in an open cluster, and more may be discovered in NGC~6791.
However, the large discrepancy between $d_{\rm YMW16}$ and $d_{\rm NGC6791}$ suggests that a coincidental spatial overlap is also a possibility.
Considering the large uncertainty of distance estimates from electron density models, it is vital to check the association with further evidence. 

There are several ways to check if PSR~J1922+37 is actually associated with NGC~6791. 
First, a precise measurement of the distance to the pulsar and the open cluster will give a straightforward answer to the problem.
The recent Gaia data release has allowed a much better estimation of the distance to NGC~6791~\citep[e.g.][]{DMM+21, HR24}, and the estimate will likely improve in future data releases.
However, the low flux density of PSR~J1922+37 will pose a challenge in measuring its distance with long baseline interferometry. No pulsar with flux density $< 10~\mu$Jy has ever had its parallax determined by interferometry.
Long-period pulsars also exhibit significant
timing noise that makes timing parallaxes
impossible, so without a long baseline array that includes FAST it is difficult to see the pulsar's distance being measurable in the near future.
A distance constraint through HI absorption measurements might be possible.

Secondly, if PSR~J1922+37 is associated with NGC~6791, then both objects should have similar proper motions.
\cite{BPC+06} reported a proper motion of $(\mu_\alpha\cos\delta, \mu_\delta)=(-0.57, -2.45)$~mas~yr$^{-1}$ for NGC~6791, while \cite{HR24} gave $(-0.42, -2.28)$~mas~yr$^{-1}$.
A measurement of the proper motion of PSR~J1922+37, e.g. via pulsar timing, would be helpful in determining the physical relation between the two objects, but timing proper motions on long-period pulsars are 
extremely difficult on anything except
decadal timescales.

Finally, like the pulsars found in globular clusters, if more pulsars with a similar DM can be found in the direction of NGC~6791, then there is a high probability that PSR~J1922+37 is associated with the open cluster.

\subsection{Formation mechanism}
\label{NGC_6791_pars}

NGC~6791 is an unusually old and massive open cluster in our Galaxy \citep{GCH+08}, with an age of several Gyr \citep{DMM+21, HR24} and a total mass ranging from 5000~M$_\odot$ \citep{PCK+11} to a few $10^4$~M$_\odot$ \citep{HR24}.
The cluster is relatively compact and follows a single-mass King density profile up to $\sim600\arcsec$, with a core region of $r_{\rm c}\sim2\farcm66$, compactness of $c\sim1.06$ and a central surface brightness of $\mu\sim17.5$ mag per arcsec$^2$ \citep{DMC+15}.
The relatively high surface brightness and compactness thus lead to a crowded stellar environment, making NGC~6791 a possible birthplace of pulsars, similar to globular clusters \citep{VER02, HCT10, BHS+13}.
If J1922+37 is associated with NGC~6791, it then becomes interesting to understand its formation mechanism.

If the pulsar is a typical slow pulsar, then in order to be associated with the cluster the pulsar must have been born in the initial wave of star formation many Gyr ago, have been
retained by the cluster despite any natal kick, and still be
an active pulsar, and be beamed at the Sun. This is improbable. $10^4$ M$_\odot$ clusters
only produce of order 50 neutron stars, and due to the low
escape velocities (a few km s$^{-1}$) most will escape the
open cluster potential. Some neutron stars may be retained, but even
in the more dense globular clusters this is only expected
to be about 10\%, so we might only expect a few percent 
in an open
cluster. Of these that are retained, most will die as radio pulsars within a few Myr. There
is a remote possibility that some of these pulsars will
have very low magnetic fields (i.e. $\sim 10^{11}$G), and might still be active
today as their spin-down rates are very small. If one is
still active, it still has to be beamed toward the Earth
to be detectable. Slow pulsars tend to have small beaming
fractions, so it is also quite unlikely. With a year of
pulsar timing the position and period derivative
should be measureable. If comparable to the age
of the cluster (i.e., many Gyr) it would strengthen the case for
the pulsar's association. We therefore recommend
a timing campaign be commenced on the pulsar.

Alternatively, the pulsar could come from a binary system, e.g. a low-mass X-ray binary, and was partially recycled before the binary was disrupted in the dynamic encounter with a cluster member \citep[e.g.][]{VF14}. 
The scenario was used by \cite{ZWL+24} to explain the slow pulsars M15K and M15L in the core region of the globular cluster M15 \citep{WPQ+24, ZWL+24}, although \cite{KYH+24} argued that star clusters are not dense enough to effectively disrupt such binaries before they are spun up to milliseconds.
% However, instead of being disrupted by an encounter, the binary system could also be destroyed by a supernova from its companion.
% In this scenario, the impact of the natal kick is reduced by the binary gravitational bound.
% This mechanism was used to interpret the formation of the magnetar CXO~J164710.2$-$455216 in Westerlund~1 \citep{BTa08, CRN+14}.

In addition to disrupted binaries, electron-captured supernovae (ECS) are also a possible pathway to producing slow pulsars in star clusters \citep{IHR+08, KYH+24, SDD+24}.
Such supernovae could be triggered by the collapse of massive white dwarfs after accretion or merger.
Since ECS is usually less energetic than core-collapse supernovae, pulsars formed in this way may suffer smaller kicks thus retain in the cluster \citep{LFR+12}.
Note that for the merger of white dwarfs, the systems could come from isolated binaries, and the merger could happen even in the case of low encounter rate.

Given the limited number of slow pulsars in star clusters, it is difficult to distinguish the various formation scenarios for this species.
The scarcity of such pulsars could partly be attributed to a bias caused by search methods; the frequently adopted FFT method is less sensitive to long-period pulsars than to millisecond ones \citep[e.g.][]{LBH+15}.
Adopting a more advanced search method, e.g. the fast folding algorithm \citep{MBS+20, ZWL+24}, will likely discover more pulsars, including the slow ones, in both star clusters and test different formation mechanisms.
We also provide a simple estimate of the expected number of recycled pulsars that could be hosted by NGC 6791 in Appendix \ref{sec:app1}.

\section{Conclusions}
\label{sec:summary}
Pulsars located or born in open clusters have not drawn much attention in pulsar research, but may hold important clues in understanding the formation and evolution of different species of pulsars and star clusters.
In this paper, we presented initial results from 20 hours of pulsar search observations in seven open clusters using FAST. 
We discovered a new pulsar J1922+37 in the direction of open cluster NGC~6791 (Fig.~\ref{fig_NGC6791}); it has a period of 1.92 second and a DM of $\sim85$~pc~cm$^{-3}$.
The NE2001 model gives a distance estimate of 4.79~kpc to the pulsar, consistent with the cluster distance, which ranges from 4.02 to 4.93~kpc in the literature.
The pulsar is thus a plausible member of NGC~6791.
If confirmed, PSR~J1922+37 will be the first pulsar discovered in an open cluster.

Measurements of the pulsar proper motion will be crucial in determining whether PSR J1922+37 is associated with NGC 6791 or not but will be difficult.
Assuming that old and massive open clusters such as NGC 6791 are similar to the Galactic globular clusters as pulsar birthplaces (e.g., the size of pulsar population is proportional to the stellar encounter rate), we estimate that around 9 recycled pulsars could be discovered in NGC 6791.
We also highlight 10 open clusters that are promising sites for finding new pulsars (in Appendix \ref{sec:app2}).

Follow-up observations have been planned with FAST, which could allow measurements of spin and astrometric parameters of PSR~J1922+37, HI absorption, as well as a deeper search for more pulsars in NGC~6791.
More pulsar discoveries with a similar DM to PSR~J1922+37 in the region would provide compelling evidence that there is a population of detectable pulsars in open clusters. If so, 
their characteristic ages could be compared with that of
the open cluster to learn how realistic pulsar characteristic
ages are, and how long slow pulsars can remain detectable
as radio pulsars.

\begin{acknowledgments}
We thank Lei Zhang, Pei Wang, Zhichen Pan, Lei Qian, Zu-Cheng Chen and Zhi-Qiang You for helpful discussions.
This work made use of the data from FAST (Five-hundred-meter Aperture Spherical radio Telescope) (\url{https://cstr.cn/31116.02.FAST}). FAST is a Chinese national mega-science facility, operated by National Astronomical Observatories, Chinese Academy of Sciences.
The data processing was mainly performed on the OzSTAR national facility at Swinburne University of Technology. 
The OzSTAR program receives funding in part from the Astronomy National Collaborative Research Infrastructure Strategy (NCRIS) allocation provided by the Australian Government, and from the Victorian Higher Education State Investment Fund (VHESIF) provided by the Victorian Government.
R.S. was supported by the National Science Foundation (NSF) grant AST-1816904.
R.P.E. is supported by the Chinese Academy of Sciences President's International Fellowship Initiative, grant No. 2021FSM0004.
JPY, NW, WWZ are supported by the National Natural Science Foundation of China (grant No. 12041304, 12041303). WWZ is supported by National SKA Program of China (grant No. 2020SKA0120200).
XJZ is supported by the National Natural Science Foundation of China (Grant No.~12203004) and by the Fundamental Research Funds for the Central Universities.

\end{acknowledgments}

\vspace{5mm}
\facilities{FAST,
            OzStar}

\software{astropy \citep{AST+13, AST+18},
          pdmp \citep{HVM04},
          peasoup \citep{BAR+20}
          ,
          psrchive \citep{HVM04}
          }
\bibliography{main}{}
\bibliographystyle{aasjournal}

\appendix

\section{Expected number of recycled pulsars in NGC~6791}
\label{sec:app1}

Following \cite{BHS+13}, we estimate the expected number of recycled pulsar discoveries, $\lambda$, in NGC~6791 in three steps.
First, using the cluster parameters in Section~\ref{NGC_6791_pars}, we computed the central luminosity density, $\rho_{\rm c}$, according to the prescription given by \cite{DJO93}. 
Second, we obtained the encounter rate $\Gamma\sim\rho_{\rm c}^2r_{\rm c}^3/\sigma_v \approx 162$ assuming a King potential model \citep{DMC+15} and a velocity dispersion of $\sigma_v\sim 2$~km~s$^{-1}$.
Note that $\Gamma$ has been normalized so that the encounter rate of 47~Tuc is $\Gamma_{\rm 47~Tuc}=1000$ \citep{BHS+13}. 
Finally, the expected number of pulsars is found by $\ln \lambda = -1.1 + 1.5\log_{10}\Gamma$ \citep{TL13, GSW+24}, which gives $\lambda\sim9$ potential discoveries.
This result is also in general agreement with the estimation using escape velocity; see \citet[][Fig. 2]{YZQ+24}. 

\section{Potential pulsar discoveries in other open clusters}
\label{sec:app2}

The possible association between PSR J1922+37 and NGC 6791 motivates us to check if there are other open clusters that may also host pulsars.
A detailed analysis inspecting the association between all known open clusters and pulsars is beyond the scope of this work, and will be presented in Zhou et al. (2024, in preparation).
Here, we check the latest Galactic cluster catalog \citep{HR24} and highlight the open clusters that are worthy of more careful pulsar surveys.

We searched the catalog and picked out the open clusters that are similar to NGC~6791 in both total mass ($>10^4$~M$_\odot$) and core size ($<10$~pc), selecting open clusters with comparable core densities and thus encounter rate \citep[e.g.][]{BHS+13}.
We also set a lower age limit of 100~Myr, which is a typical lifetime for canonical pulsars.

Ten open clusters, including NGC~6791, meet the above conditions and are shown in Table~\ref{tab:oc_list}, which lists the total mass, angular size, and number of member stars.
All the clusters are relatively close ($d<5$~kpc) and have a small core size with a typical value of $\sim 5\arcmin$.
They also have a relatively high stellar concentration, with half of members within the tidal radius.
These clusters are thus good candidates for future pulsar searches in open clusters.

\begin{table*}
    \centering
\begin{tabular}{ccrcccccccccc}
\toprule
Name  &    RA     &  Dec      &   $d$     &  Age          &  $M_{\rm total}$ & $\sigma_{V_{\rm r}}$  &  \multicolumn{2}{c}{$r_{\rm c}$}   &    $r_{\rm tidal}$      &  $r_{\rm total}$     & $N_{\rm tidal}$  &    $N_{\rm total}$  \\ 
    &   (deg)   &  (deg)    & (kpc)   & ($10^8$~yr)    & ($10^4$~M$_{\odot}$)    & (km~s$^{-1}$)  & (pc) & (min) & (min)   & (min)  &         &       \\
    \hline
Berkeley 53  &  313.98   &  51.05    &  3.41   &   5.6         &  3.73            &   0.6                 &  3.9  & 3.9      &   11.4     &   26.7    &    714  & 1660  \\     
Theia 1661  &  267.59   & $-25.15$  &  2.30   &   2.4         &  2.71            &   2.6                 &  2.6  & 3.9      &  105.5     &  113.9    &   2607  & 2620  \\     
Trumpler 5  &   99.13   &   9.46    &  2.90   &  21.7         &  2.46            &   0.6                 &  3.9  & 4.6      &   30.2     &   44.2    &   2551  & 3619  \\     
NGC 2158  &   91.86   &  24.10    &  3.71   &  18.6         &  2.09            &   1.3                 &  2.0  & 1.8      &   12.6     &   39.6    &   1207  & 1958  \\     
IC 166  &   28.09   &  61.86    &  4.42   &   7.8         &  2.07            &   0.7                 &  2.7  & 2.1      &   11.5     &   28.5    &    801  & 1768  \\
\\
NGC 6791  &  290.22   &  37.77    &  4.19   &  38.0         &  1.95            &   2.2                 &  4.5  & 3.7      &   11.6     &   41.5    &   2062  & 4193 \\     
NGC 7789  &  359.33   &  56.72    &  1.97   &  15.7         &  1.78            &   0.9                 &  4.0  & 6.9      &   32.2     &   49.5    &   2377  & 4125  \\     
NGC 6259  &  255.19   & $-44.68$  &  2.16   &   1.6         &  1.78            &   1.8                 &  4.1  & 6.6      &   23.4     &   34.7    &   1310  & 2188  \\     
Pismis 3  &  127.83   & $-38.66$  &  2.11   &  10.6         &  1.59            &   0.8                 &  3.6  & 5.8      &   22.0     &   68.1    &   1204  & 2879  \\     
Trumpler 20  &  189.89   & $-60.63$  &  3.28   &  15.8         &  1.53            &   0.5                 &  9.7  &10.2      &   10.2     &   53.4    &    811  & 1692  \\     
\hline
\end{tabular}
    \caption{Open clusters similar to NGC~6791 in order of decreasing total mass.
    Columns are cluster name, right ascension and declination, distance, age, total mass, the standard error of radial velocity, core radius (in pc and in arcmin), tidal radius, total radius, number of member stars within $r_{\rm tidal}$ and the total number of stars.
    The parameters of NGC~6791 are also included for comparison.
    }
    \label{tab:oc_list}
\end{table*}

\end{document}